\documentstyle[preprint,tighten,aps]{revtex}
\begin{document}
\draft
\baselineskip=1.05\baselineskip
\title
{Five-loop renormalization-group expansions \\ for the three-dimensional 
$n$-vector cubic model \\ and critical exponents for impure Ising systems}

\author{D.~V.~Pakhnin$^1$ and A.~I.~Sokolov$^{1,2}$}

\address
{$^1$Department of Physical Electronics, 
Saint Petersburg Electrotechnical University, \\ Professor Popov Street 5,
Saint Petersburg 197376, Russia, \\
$^2$Department of Physics, Saint Petersburg Electrotechnical University, \\
Professor Popov Street 5, Saint Petersburg 197376, Russia}

\maketitle

\begin{abstract}
The renormalization-group (RG) functions for the three-dimensional 
$n$-vector cubic model are calculated in the five-loop approximation. 
High-precision numerical estimates for the asymptotic critical 
exponents of the three-dimensional impure Ising systems are extracted 
from the five-loop RG series by means of the Pad\'e-Borel-Leroy 
resummation under $n = 0$. These exponents are found to be: 
$\gamma = 1.325 \pm 0.003$, ~$\eta = 0.025 \pm 0.01$, 
~$\nu = 0.671 \pm 0.005$, ~$\alpha = - 0.0125 \pm 0.008$, 
and $\beta = 0.344 \pm 0.006$. For the correction-to-scaling exponent, 
the less accurate estimate $\omega = 0.32 \pm 0.06$ is obtained.    
\vspace{1.5cm}
 
PACS numbers: 75.40.Cx, 64.60.Ak, 11.10.Lm, 64.60.Fr 

\end{abstract}

\newpage

\section{Introduction}
\label{sec:1}

The critical thermodynamics of cubic crystals and weakly disordered 
systems has remained an area of extensive theoretical work during 
past decades. Considerable progress in studying the random critical 
behavior was achieved 25 years ago when Harris and Lubensky 
\cite{HL,TCL} and Khmelnitskii \cite{KDE}  
attacked the problem by the field-theoretical renormalization-group 
(RG) approach based on the Euclidean scalar $\varphi^4$ theory in 
$(4 - \epsilon)$ dimensions. As a results, the regular method for 
calculating critical exponents and other universal quantities of the 
impure Ising model -- the famous $\sqrt{\epsilon}$-expansion -- was 
invented. The numerical power of this technique, however, stayed for a 
long time unclear since only lower-order contributions to critical 
exponents and the equation of state have been found 
\cite{HL,TCL,KDE,BNS1,JK}. Recently, starting from the five-loop RG 
series obtained for the $(4 - \epsilon)$-dimensional cubic model by 
Kleinert and Schulte-Frohlinde \cite{HK}, the calculation of 
the $\sqrt{\epsilon}$-expansions for critical exponents was performed 
up to the $\sqrt{\epsilon}^4$ and $\sqrt{\epsilon}^5$ terms \cite{SAS}.
As was found, these series possess a rather irregular structure making 
them unsuitable for subsequent resummation and, hence, practically 
useless for getting numerical estimates \cite{FHY1}. 
      
On the other hand, there exists an alternative field-theoretical approach 
that proved to be very efficient when used for evaluation of universal 
critical quantities. We mean the perturbative renormalization group in 
three dimensions (3D) yielding most accurate numerical estimates for 
critical exponents, critical amplitude ratios, and universal higher-order 
couplings of the O(n)-symmetric systems 
\cite{BNM,LGZ,AS,N,GZ1,GZ2,S,K1,K2,K3,SOUK,SLD}. 
The impure Ising model at criticality is known to be described by the 
$n$-vector field theory with the quartic self-interaction having a 
hupercubic symmetry, provided $n \rightarrow 0$ (the replica limit) and 
the coupling constants have proper signs. In the 1980s, the RG expansions 
for three-dimensional (3D) cubic and impure Ising models have been 
calculated in the two-loop \cite{GJ}, three-loop \cite{SS,SH}, and 
four-loop \cite{MSS,IOM} approximations paving the way for estimating 
the critical exponents and other universal quantities 
\cite{GJ,SS,SH,MSS,IOM,K4,MS,MSI,SHP,BS,K5,IM,FHY,KBV}. 
The numerical results thus obtained was found to agree, in general, with 
the most accurate experimental and simulation data. 

In the course of this study, it was revealed, however, that even the 
highest-order available, four-loop 3D RG expansions, when resummed 
by means of the generalized Pad\'e-Borel-Leroy method 
do not allow us, in fact, to optimize the resummation procedure, i. e., 
to choose the best Pad\'e approximant and the optimal value of the tune 
parameter since there is the only approximant -- [3/1] -- 
that does not suffer from positive axis poles. Moreover, an account for 
four-loop terms in the 3D RG expansions shifts the fixed point coordinates 
and the value of the correction-to-scaling exponent $\omega$ appreciably 
with respect to their three-loop analogs, indicating that at this step 
the RG-based iterations do not still achieve their asymptote. 
This prevents the four-loop RG approximation from to be thought of as 
sufficient, i. e., providing, within the perturbation theory, the accurate 
theoretical predictions. 

In such a situation, a calculation of the higher-order contributions to 
the RG functions looks very desirable. In this paper, 
the five-loop RG expansions for the three-dimensional cubic model are 
obtained and resulting numerical estimates for the critical exponents 
of the weakly disordered Ising systems are found. 

\section{RG expansions for $\beta$ functions and critical exponents}
\label{sec:2}

The Landau-Wilson Hamiltonian of the three-dimensional $n$-vector 
cubic model reads:
\begin{equation}
H = {1 \over 2} 
\int d^3x \Biggl[ m_0^2 \varphi_{\alpha}^2 + (\nabla \varphi_{\alpha})^2 
+ {u_0 \over 12} \varphi_{\alpha}^2 \varphi_{\beta}^2 
+ {v_0 \over 12} \varphi_{\alpha}^4 \Biggr], 
\label{eq:1}
\end{equation}
where $\varphi$ is an $n$-component real order parameter, $m_o^2$ being 
the reduced deviation from the mean-field transition temperature.
In the replica limit, this Hamiltonian describes the critical behavior of 
the impure Ising model provided $u_o < 0$ and $v_o > 0$.

We calculate the $\beta$ functions for the Hamiltonian 
Eq.~(\ref{eq:1}) within a massive theory. The renormalized Green 
function $G_R(p,m)$, the $\varphi^2$ insertion and four-point vertices 
$U_R(p_i,m,u,v,)$, $V_R(p_i,m,u,v,)$ are normalized 
at zero external momenta in a conventional way:
\begin{eqnarray}
G_R^{-1}(0,m) = m^2, \ \ \ \  
{{\partial G_R^{-1}(p,m)} \over {\partial p^2}} 
\Big\arrowvert_{p^2 = 0} = 1, \ \ \ \
\nonumber \\
\Gamma_R^{(1,2)}(p,q,m,u,v) \Big\arrowvert_{p = q = 0} = 1, 
\quad \qquad \qquad \qquad 
\nonumber \\ 
U_R(0,m,u,v) = mu, \ \ \ \ \ V_R(0,m,u,v) = mv. \
\label{eq:2}
\end{eqnarray}
The value of the one-loop vertex graph including the factor $(n + 8)$ is
absorbed in $u$ and $v$ in order to make the coefficient for the 
$u^2$ term in $\beta_u$ equal to unity. 

The four-loop RG expansions for the functions of interest have been found 
earlier \cite{MSS}. To extend these series to the five-loop order, we   
calculate corresponding tensor (field) factors generated by the 
O(n)-symmetric and cubic interactions. Taking, then, numerical values of 
the 3D integrals from Ref.\cite{NMB}, we arrive to the following 
five-loop expansions: 
\begin{eqnarray}
{\beta_u \over u} &=& 1 - u - {6 v \over (n + 8)}
+ {4 \over {27(n + 8)^2}} \biggl[(41 n + 190) u^2 
+ 300 u v + 69 v^2 \biggr] 
\nonumber\\&&
- {1 \over (n + 8)^3} \biggl[( 1.34894276 n^2 + 54.9403770 n 
+ 199.640417 ) u^3 + ( 19.9406350 n 
\nonumber\\&&
+ 493.841548 ) u^2 v + (1.86566761 n + 302.867786) u v^2 
+ 65.9372851 v^3 \biggr] 
\nonumber\\&& 
+ {1 \over (n + 8)^4} \biggl[( - 0.155645891 n^3 + 35.8202038 n^2 
+ 602.521231 n + 1832.20673 ) u^4 
\nonumber\\&&
+ ( - 4.05864597 n^2 + 546.221669 n 
+ 6192.51210 ) u^3 v + ( 81.7510086 n 
\nonumber\\&&
+ 6331.22642 ) u^2 v^2 + ( 11.6191565 n + 2777.39424 ) u v^3 
+ 495.005747 v^4 \biggr] 
\nonumber\\&&
- {1 \over (n + 8)^5} \biggl[( 0.0512361755 n^4 + 3.23787619 n^3 
+ 668.554337 n^2 + 7819.56476 n 
\nonumber\\&&
+ 20770.17697 ) u^5 + ( 1.87656422 n^3 + 65.4181099 n^2 + 11485.34719 n 
+ 89807.66984 ) u^4 v 
\nonumber\\&&
+ ( 21.0505258 n^2 + 3858.04476 n + 130340.90533 ) u^3 v^2 
+ ( 630.460362 n 
\nonumber\\&&
+ 90437.63644 ) u^2 v^3 
+ ( 79.5359421 n + 33088.22288 ) u v^4 + 5166.39201 v^5 \biggr] \ \ , 
\label{eq:3}
\end{eqnarray}

\begin{eqnarray}
{\beta_v \over v} &=& 1 - {(12 u + 9 v) \over (n + 8)}  
+ {4 \over {27(n + 8)^2}} \biggl[(23 n + 370) u^2 
+ 624 u v + 231 v^2 \biggr] 
\nonumber\\&&
- {1 \over (n + 8)^3} \biggl[(- 1.25110731 n^2 + 41.8539021 n 
+ 469.333970 ) u^3 
\nonumber\\&&
+ ( 2.23905886 n + 1228.60591 ) u^2 v 
+ 957.781662 u v^2 + 255.929737 v^3 \biggr]  
\nonumber\\&&
+ {1 \over (n + 8)^4} \biggl[( 0.574652520 n^3 - 0.267107207 n^2 
+ 584.287672 n + 5032.69226 ) u^4 
\nonumber\\&&
+ ( 0.172125857 n^2 
+ 322.925039 n + 17967.85060 ) u^3 v 
\nonumber\\&&
+ (- 49.4820078 n + 21964.39381 ) u^2 v^2 
+ 11856.95686 u v^3 + 2470.39252 v^4 \biggr]  
\nonumber\\&&
- {1 \over (n + 8)^5} \biggl[( - 0.318104330 n^4 - 3.62982162 n^3 
+ 139.264889 n^2 + 9324.60054 n 
\nonumber\\&&
+ 64749.28195 ) u^5 
+ ( - 1.14454168 n^3 - 122.339901 n^2 + 10376.55804 n 
\nonumber\\&&
+ 294450.70368 ) u^4 v
+ ( 12.6147089 n^2 + 233.955446 n + 493917.03678 ) u^3 v^2 
\nonumber\\&&
+ ( - 1363.28787 n + 407119.30675 ) u^2 v^3 + 170403.11905 u v^4 
+ 29261.58518 v^5 \biggr].
\label{eq:4}
\end{eqnarray}

\begin{eqnarray}
\gamma^{-1} &=& 1 - {{(n + 2) u + 3 v} \over {2(n + 8)}} 
+ {1 \over (n + 8)^2} \biggl[(n + 2) u^2 + 6 u v + 3 v^2 \biggr] 
\nonumber\\&&
- {1 \over (n + 8)^3} \biggl[(0.879558892 n^2 + 6.48547686 n 
+ 9.45271816) u^3 + (7.91603003 n 
\nonumber\\&&
+ 42.5372317) u^2 v
+ (1.15505603 n + 49.2982057) u v^2 + 16.8177539 v^3 \biggr]
\nonumber\\&&
+ {1 \over (n + 8)^4} \biggl[(-0.128332104 n^3 
+ 7.96674070 n^2 + 51.8442130 n + 70.7948063) u^4 
\nonumber\\&&
+ (-1.53998525 n^2 + 98.6808589 n + 424.768838) u^3 v 
+ (30.8151755 n 
\nonumber\\&&
+ 752.049392) u^2 v^2  
+ (5.64296122 n + 516.266750) u v^3 + 130.477428 v^4 \biggr]
\nonumber\\&&
- {1 \over (n + 8)^5} \biggl[(0.0490966055 n^4 + 4.28815249 n^3 
+ 108.361822 n^2 + 537.813610 n 
\nonumber\\&&
+ 675.699608) u^5  + (0.736449089 n^3 + 62.8493893 n^2 
+ 1499.72855 n + 5067.74706) u^4 v 
\nonumber\\&&
+ (9.54286409 n^2 + 1059.53469 n + 12193.04534) u^3 v^2 + (295.911053 n 
\nonumber\\&&
+ 12966.21184) u^2 v^3 + (43.2275845 n + 6587.83386) u v^4 
+ 1326.21229 v^5 \biggr]
\label{eq:5}
\end{eqnarray}

\begin{eqnarray}
\eta &=& {8 \over {27 (n + 8)^2}} \biggl[(n + 2) u^2 + 6 u v 
+ 3 v^2 \biggr]  
+ {1 \over (n + 8)^3} \biggl[( 0.0246840014 (n^2 + 10 n + 16) u^3 
\nonumber\\&&
+ 0.222156013 (n + 8) u^2 v + 1.99940412 u v^2 + 0.666468039 v^3 \biggr] 
\nonumber\\&&
+ {1 \over (n + 8)^4} \biggl[(-0.0042985626 n^3 + 0.667985921 n^2 
+ 4.60922106 n + 6.51210994) u^4 
\nonumber\\&&
+ (-0.0515827507 n^2 + 8.118996555 n + 39.0726596) u^3 v 
\nonumber\\&&
+ (2.04484677 n + 68.6652634) u^2 v^2   
+ 47.1400734 u v^3 + 11.7850184 v^4 \biggr] 
\nonumber\\&&
- {1 \over (n + 8)^5} \biggl[(0.00655092214 n^4 - 0.132451063 n^3 
+ 1.89113927 n^2 
\nonumber\\&&
+ 15.1880934 n + 21.6472064) u^5 
+ (0.0982638321 n^3 - 2.18329361 n^2 + 32.7336762 n 
\nonumber\\&&
+ 162.354048) u^4 v + (-0.164297296 n^2 + 4.14587079 n 
+ 382.023815) u^3 v^2 
\nonumber\\&&
+ (-3.99131884 n + 389.996707) u^2 v^3 + 193.002694 u v^4 
+ 38.6005389 v^5 \biggr] 
\label{eq:6}
\end{eqnarray}
These expansions will be used to evaluate the critical exponents 
of the impure Ising model.

\section{Resummation and numerical estimates}
\label{sec:3}

Numerical values of critical exponents are known to be determined 
by the coordinates of a relevant fixed point. In our case, i. e., for 
$n = 0$, a point of interest is the random fixed point. To find its 
location and, more generally, to extract the physical information from 
divergent RG series, a proper resummation procedure should be applied.  
Here, we use the Pad\'e-Borel-Leroy resummation technique, which 
demonstrates high numerical effectiveness both for the O(n)-symmetric 
models \cite{BNM,AS,S} and for anisotropic systems preserving their 
internal symmetries (see, e. g. Ref.{\cite{AS1}) for detail). Since the 
expansions of quantities depending on two variables $u$ and $v$ are dealt 
with, the Borel-Leroy transformation is taken in a generalized form: 
\begin{equation}
f(u,v)=\sum_{ij}c_{ij}u^iv^j=\int\limits_0^\infty
e^{-t} t^b F(ut,vt)dt, \qquad
F(x,y)=\sum_{ij}{\frac{c_{ij}x^i y^j}{{(i+j+b)!}}}. 
\label{eq:7}
\end{equation}
To perform an analytical continuation, the resolvent series
\begin{equation}
\tilde F(x,y,\lambda )=\sum_{n=0}^\infty \lambda
^n\sum_{l=0}^n{\frac{c_{l,n-l}x^ly^{n-l}}{{(n+b)!}}}
\label{eq:8}
\end{equation}
is constructed, which is a series in powers of $\lambda$ with 
coefficients being uniform polynomials in $x$, $y$ and then Pad\'e
approximants $[L/M]$ in $\lambda $ at $\lambda = 1$ are used. 

For the resummation of the five-loop RG expansions, we employ three 
different Pad\'e approximants: [4/1], [3/2], and [2/3]. 
The first of them, being pole-free, is known to give good numerical 
results for 3D O(n)-symmetric models while the others are near-diagonal 
and should reveal, {\em a priory}, the best approximating properties. 
The coordinates of the random fixed point resulting from the series 
Eqs.~(\ref{eq:3}, \ref{eq:4}) resummed using these approximants 
under $b = 0$ and $b = 1$ are presented in Table I, which also contains  
analogous estimates given by the four-loop RG expansions. 
The four-loop series were processed on the base of the Pad\'e 
approximant [3/1], since use of the diagonal approximant [2/2] leads 
to the integrand in Eq.~(\ref{eq:7}) that 
has a dangerous pole in the vicinity of the random fixed point both 
for $\beta_u$ and $\beta_v$ \cite{32}. The fixed-point location given 
by the approximant [2/3] is presented for $b = 0$ only, because for 
$b = 1$ this approximation predicts no random fixed point.  
 
As is seen from Table I, Pad\'e approximants [4/1] and 
[3/2] yield numerical values of the random fixed-point coordinates, which 
are remarkably close to each other. Moreover, for $b = 0$ they are also 
close to those given by the approximant [3/1]: the largest difference 
between the five-loop and four-loop estimates does not exceed 0.026. 
With increasing $b$, corresponding numbers diverge, indicating that 
$b = 0$ is an optimal value of the tune parameter. On the contrary, 
Pad\'e approximant [2/3] gives the random fixed-point 
location, which deviates appreciably from those predicted by approximants 
[4/1], [3/2], and [3/1]. This approximant, however, is found to lead 
to poor numerical results even for simpler systems, e. g., for the Ising 
model. Indeed, when used to evaluate the coordinate of the Ising fixed 
point, it results in $v_c = 1.475$ (under $b = 0$) while the best  
estimate today is known to be $v_c = 1.411$ \cite{GZ2}. This forces 
us to reject the data obtained on the base of the approximant [2/3].

So, to determine the random fixed-point coordinates, we have to average 
the numerical data given by three working Pad\'e 
approximants at $b = 0$. This procedure yields the values  
\begin{equation}
u_c = -0.711, \qquad \qquad
v_c = 2.008, \quad \quad
\label{eq:9}
\end{equation}
which are claimed to be the final results of our search of the 
random fixed-point location. Let us estimate their accuracy. 
It seems unlikely that the deviations
of these numbers from the exact ones would exceed the differences 
between them and the four-loop estimates since, among all proper 
estimates, the four-loop ones most strongly differ from the averaged 
values (\ref{eq:9}). Hence, the error bounds for $u_c$ and $v_c$ are 
believed to be not greater than $\pm 0.012$ and $\pm 0.016$, 
respectively. 
Another way to estimate an apparent accuracy is to trace how the 
averaged values of the random fixed point coordinates vary with the 
variation of $b$. We calculate $u_c$ and $v_c$ using the pole-free 
approximants [4/1] and [3/1] for $0 \le b \le 15$; $b = 15$ 
is chosen as a largest reasonable value of the tune parameter, since 
for greater $b$ the saturation of dependences of various quantities 
on $b$ becomes visible. Running through this interval, the averaged 
coordinates change their values by about 0.02 indicating that an 
accuracy of the estimates (\ref{eq:9}) is of the order of 0.01-0.02,  
in accord with that found above. 

With the numbers (\ref{eq:9}) in hand, we can evaluate the critical 
exponents for the 3D impure Ising model. The exponent $\gamma$ is 
estimated by the Pad\'e-Borel-Leroy resummation of the RG series 
(\ref{eq:5}) for $\gamma^{-1}$ and of the inverse series, i. e., the 
RG expansion for $\gamma$. The Fisher exponent is also evaluated in 
two different ways: via the estimation of the critical exponent 
$\eta_2 = (2 - \eta)(\gamma^{-1} - 1)$ having the RG expansion, 
which exhibits a good summability, and by direct substitution of the 
fixed point coordinates into the series (\ref{eq:6}) with rapidly 
diminishing coefficients. The estimates for $\eta$ originating from 
$\eta_2$ were obtained under the central value of $\gamma$: 
$\gamma = 1.325$. The numerical results thus found are collected 
in Table II. 

As is seen from this Table, two methods of evaluating of the 
susceptibility exponent $\gamma$ lead to remarkably close numerical 
results, which very weakly depend on the tune parameter. 
Indeed, with increasing $b$ from 0 to 15, the estimates 
for $\gamma$ obtained by the resummation of the RG 
series for $\gamma$ and $\gamma^{-1}$ on the base of the pole-free 
approximant [4/1] vary by less than 0.0036, while the difference between 
them never exceeds 0.0013. Under the same variation of $b$, the value of 
$\gamma$ averaged over these two most reliable approximations remains 
within the segment [1.3240, 1.3266]. On the other hand, the accuracy of 
determination of the critical exponents depends not only on a quality 
of the resummation procedure but also on the accuracy achieved in the 
course of locating of the relevant fixed point. That is why we 
investigated to what extent the numerical estimates for $\gamma$ vary 
when coordinates of the random fixed point run through their error bars. 
It was found that the susceptibility exponent calculated 
at the optimal value of tune parameter $b = 2$ (see Table II) does not 
leave the segment [1.3228, 1.3263]. This enables us to conclude that 
the error bounds for the theoretical value of $\gamma$ obtained in this 
work would be about $\pm 0.003$ or smaller. 

Less stable numerical results, with respect to a variation of $b$, 
are found for the Fisher exponent $\eta$. As one can see from Table II,
the values of $\eta$ given by the resummation of the RG series for 
$\eta_2$ with use of the pole-free Pad\'e approximant [4/1] spread 
from 0.0148 to 0.0312. The average over this interval is equal to 0.023. 
The direct summation of the RG expansion for $\eta$ at the random fixed 
point gives 0.027. It is natural to suppose that 0.025 should play a 
role of the most likely value of exponent $\eta$. Since the estimates 
for $\eta$ found via the evaluation of $\eta_2$ are sensitive to 
the accepted value of $\gamma$, the apparent accuracy achieved in this 
case is not believed to be better than $\pm 0.01$. 

Having estimated $\gamma$ and $\eta$, we can evaluate other critical 
exponents using well-known scaling relations. So, the final results of 
our five-loop RG analysis are as follows:
\begin{eqnarray}
\gamma = 1.325 \pm 0.003, \qquad 
\eta = 0.025 \pm 0.01, \qquad 
\nu = 0.671 \pm 0.005, 
\nonumber \\ 
\alpha = - 0.0125 \pm 0.008, \qquad 
\beta = 0.344 \pm 0.006. \qquad \qquad \qquad  
\label{eq:10}
\end{eqnarray}
These numbers are thought to be the most accurate theoretical estimates 
for the critical exponents of the 3D impure Ising model known today. 
It is interesting to compare them with those obtained earlier
within the lower-order RG approximations. For the exponent $\gamma$,
previous RG calculations in three dimensions gave the values 
1.337 \cite{GJ,MS} (two-loop), 1.328 \cite{SH} (three-loop), 1.326 
\cite{MSS} (four-loop), and 1.321 \cite{IOM} (four-loop). Being found 
by means of the different resummation procedures, they are, nevertheless, 
centered around our estimate which is thus argued to be very close to 
the exact value of $\gamma$ or, more precisely, to the true asymptote 
of the RG iterations.   

At the end of this work, we employ our technique to evaluate 
the correction-to-scaling exponent $\omega$. 
The exponent $\omega$ is known to be equal to the stability matrix 
eigenvalue that has a minimal modulus. The derivatives 
$\partial \beta_u/ \partial u$, $\partial \beta_u/ \partial v$, 
$\partial \beta_v/ \partial u$, and $\partial \beta_v/ \partial v$ 
entering this matrix are evaluated numerically at the random fixed 
point on the base of the resummed RG expansions for $\beta_u$ and 
$\beta_v$, and then the matrix eigenvalues are found. 
Such a procedure leads to the estimates for $\omega$ presented in 
Table I (lower lines). They are seen to be considerably scattered 
and sensitive to the tune parameter. The average over three working 
Pad\'e approximants, however, being equal to 0.315 at $b = 0$ and 
to 0.316 at $b = 1$ turns out to be stable under the variation of 
$b$ unless $b$ becomes large. It is natural therefore to accept that 
\begin{equation}
\omega = 0.32 \pm 0.06.
\label{eq:11}
\end{equation}
This number is smaller by $0.05-0.07$ than its counterparts given 
by recent Monte Carlo simulations \cite{B} and the alternative RG 
analysis \cite{FHY}, but their central values lie within the declared 
error bounds (\ref{eq:11}). Hence, an agreement between the results 
discussed exists. On the other hand, because of its low accuracy, 
the estimate (\ref{eq:11}) would not be thought of as satisfactory. 
It certainly needs to be improved, along with the estimates for the 
small exponents $\eta$ and $\alpha$ also exhibiting appreciable 
uncertainties. The only way to do this is the studying of the 
critical behavior of the 3D impure Ising model in the next 
perturbative order that includes calculations of the six-loop terms 
in the relevant RG expansions. Such calculations are now in progress.    

\section{Conclusion}    
\label{sec:4}
    
To summarize, we have calculated the five-loop RG expansions for the   
$\beta$ functions and critical exponents of the 3D $n$-vector cubic 
model. The resummation of the RG series by the Pad\'e-Borel-Leroy 
technique in the replica limit ($n=0$) has enabled us to obtain 
high-precision numerical estimates for the random fixed-point 
coordinates and the critical exponents $\gamma$, $\eta$, $\alpha$, 
$\nu$, and $\beta$ of the 3D impure Ising model. For the 
correstion-to-scaling exponent $\omega$, the resummed five-loop RG 
expansions turned out to give much less accurate numerical results. 
The values of the critical exponents obtained earlier from the 
lower-order RG expansions in three dimensions are shown to be 
centered by our five-loop estimates. This indicates that the five-loop 
RG approximation provides numerical data very close to the asymptotic 
ones, i. e., to those representing the point of convergence of the 
RG-based iterations. At the same time, an accuracy achieved when 
evaluating the exponent $\omega$ would not be referred to as sufficient 
making the next-order, six-loop RG calculations, very desirable.     

{\em Note added}. After this paper had been submitted for publication, 
Ref.{\cite{CPV}} appeared where the six-loop 
RG expansions for the 3D cubic model are calculated. 
The five-loop terms obtained in this remarkable work are found 
to agree with ours. Among other quantities, in Ref.{\cite{CPV}} the 
marginal value of $n$, $n_c$, separating different regimes of critical 
behavior of the cibic model, is estimated. To clear up how sensitive to 
the method of resummation the five-loop RG results are, we evaluate 
$n_c$ using Pad\'e-Borel-Leroy technique. Exploiting the near-diagonal
Pad\'e approximant [3/2] under $b$ varying from 0 to 20, the values of 
$n_c$ are obtained which lie between 2.89 and 2.92. They agree quite 
well with the estimate $n_c = 2.91(3)$ extracted from the five-loop RG
series by means of the conformal-mapping-based resummation machinery 
\cite{CPV}.

\acknowledgments 

This work was supported by the Ministry of Education of Russian 
Federation under Grant No. 97-14.2-16. One of the authors (A.I.S.) 
gratefully acknowledges also the support of the International Science 
Foundation via Grant p99--943.

\narrowtext
\begin{table}
\caption{Numerical estimates for the random fixed-point location and 
correction-to-scaling exponent $\omega$ obtained from the five-loop RG 
expansions (\ref{eq:3}, \ref{eq:4}) resummed by the Pad\'e-Borel-Leroy 
technique using approximants $[4/1]$, $[3/2]$, and $[2/3]$. The last 
column contains results given by the four-loop RG series processed on 
the base of the approximant $[3/1]$. The superscript "c" denotes that 
the exponent $\omega$ is complex and its real part is presented. The 
superscript "p" stands to mark that the Pad\'e approximant has 
a "nondangerous" positive axis pole, i. e. a pole well remoted from 
the origin ($t > 40$) that affects neither the procedure of numerical 
evaluation nor the value itself of the Borel integral.}
\begin{tabular}{cccccc}
& b & $[4/1]$ & $[3/2]$ & $[2/3]$ & $[3/1]$ \\
\tableline
$u_c$ & 0 & $-0.7200$ & $-0.7148$ & $-0.6871$ & $-0.6991$  \\
& 1 & $-0.7445$ & $-0.7385^p$ & & $-0.6839$ \\
\tableline
$v_c$ & 0 & 2.0182 & 2.0125 & 2.0571 & 1.9922 \\
& 1 & 2.0296 & $2.0236^p$ & & 1.9877 \\
\tableline
$\omega$ & 0 & 0.266 & 0.303 & $0.462^c$ & 0.376 \\
& 1 & 0.263 & $0.325^p$ & & 0.361 \\
\end{tabular}
\label{table1}
\end{table}

\begin{table}
\caption{Numerical estimates for the critical exponents $\gamma$ and 
$\eta$ obtained from the five-loop RG expansions (\ref{eq:5}, \ref{eq:6}) 
resummed by the Pad$\acute {{\rm e}}$-Borel-Leroy technique using 
approximants $[4/1]$ and $[3/2]$. DS stands for "direct summation", 
the symbol $(\gamma^{-1})^{-1}$ means that the RG series for 
$\gamma^{-1}$ was resummed. The superscript "p" denotes, as in Table 1, 
that the Pad\'e approximant has a "nondangerous" pole, while empty 
cells are due to the dangerous ones spoiling corresponding 
approximations. The estimates for $\eta$ standing in the fifth and 
sixth lines were obtained under $\gamma = 1.325$ by the resummation 
of the RG series for $\eta_2$.}
\begin{tabular}{cccccccccc}
b &   & 0 & 1 & 2 & 3 & 4 & 5 & 10 & 15 \\
\tableline
$(\gamma^{-1})^{-1}$ & [4/1] & 1.3236 & 1.3244 & 1.3250 & 1.3254 
& 1.3257 & 1.3260 & 1.3268 & 1.3272 \\
& [3/2] & - & - & - & $1.3253^p$ & $1.3257^p$ & 1.3260 & 1.3265 & 1.3267 \\
$\gamma$  & [4/1] & 1.3245 & 1.3248 & 1.3250 & 1.3252 
& 1.3253 & 1.3254 & 1.3257 & 1.3259 \\
& [3/2] & $1.3246^p$ & $1.3251^p$ & $1.3254^p$ & $1.3257^p$ 
& $1.3259^p$ & $1.3261^p$ & $1.3267^p$ & $1.3270^p$ \\
\tableline
$\eta$ (via $\eta_2$) & [4/1] & 0.03122 & 0.02765 & 0.02506 & 0.02311 
& 0.02158 & 0.02036 & 0.01665 & 0.01478 \\
& [3/2] & - & - & - & $0.02870^p$ & $0.02419^p$ 
& $0.02168^p$ & $0.01666^p$ & 0.01487 \\
$\eta$ (DS) &  &  &  &  & 0.0272 &  &  &  & \\
\tableline
\end{tabular}
\label{table2}
\end{table}

\begin{references}
\bibitem{32} In fact, under $b = 1$ the approximant [3/2] generates the 
expression for $\beta_v$ that is also spoilt by a positive axis pole at 
the random fixed-point location. This pole, however, being well remoted  
from the origin ($t = 40.12$), turns out to be "nondangerous", i. e., 
does not influence, in practice, upon the evaluation of the Borel 
integral.
\end{references}
\end{document}